\documentclass[12pt]{article}

\usepackage{times}
\usepackage[pdftex]{graphicx}
\usepackage[figuresright]{rotating}

\begin{document}

\thispagestyle{empty}       
\title{Highlights from Heavy Ion Collisions at RHIC and the Acoustics of the Little Bangs }

\author{Paul Richard Sorensen\\
{\small \it Brookhaven National Laboratory, Upton, New York 11973, USA }\\
}
\date{}

\thispagestyle{empty}       
\maketitle
\thispagestyle{empty}       
\pagestyle{empty}

\begin{abstract}
  At the Relativistic Heavy Ion Collider (RHIC) at Brookhaven National
  Laboratory, heavy nuclei are collided at high energies to
  create matter that is hot enough and dense enough to dissolve hadrons into
  a quark-gluon-plasma (QGP). In this lecture, dedicated to the memory
  of Aditya Sambamurti, I present an introduction to heavy-ion
  collisions and highlights from the first decade of RHIC results.
\end{abstract}

\section{Introduction}
The motivation for the construction of the Relativistic Heavy Ion
Collider at Brookhaven National Laboratory was to collide heavy
nuclei so that a theoretically conjectured state of matter
called the Quark Gluon
Plasma (QGP)~\cite{Reisdorf:1997fx,Herrmann:1999wu} could be formed. These collisions
deposit relatively large amounts of energy into a region the size
of a nucleus. The matter left behind in that region is so hot and
dense that hadronic matter undergoes a phase transition into a form of
matter where quarks and gluons are the relevant degrees of freedom,
not hadrons~\cite{eos}. This is the state of matter that existed when
the universe was young (less than a microsecond old) and very hot.

\begin{figure}[htb]
  \centering
  \resizebox{0.65\textwidth}{!}{\includegraphics{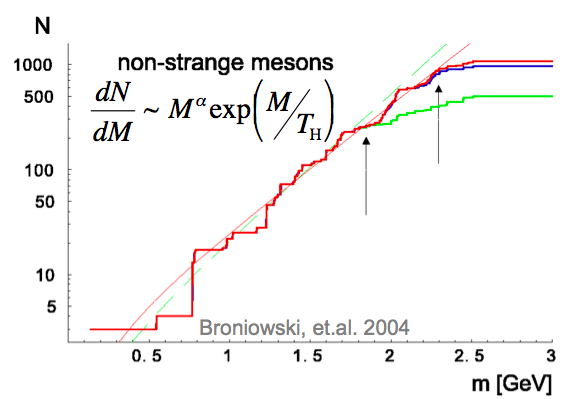}}
  \caption[]{ The hadronic mass spectrum; the number of hadronic
    states increases exponentially with mass. This exponential
    increase implies a maximum temperature for hadronic matter called
    the Hagedorn temperature. Figure from Ref.~\cite{Broniowski:2004yh} }
\label{f1}
\end{figure}

The first clues to the existence of a transition from hadronic matter
to a different state of matter predate the discovery of quarks and the
advent of Quantum Chromodynamics. Even before the quark model was
proposed, Hagedorn noticed that the spectrum of hadronic states (shown
in figure~\ref{f1}) grows exponentially with mass; that is, the number
of hadronic states within a given mass window, increases exponentially with the
mass. Hagedorn realized that this feature of the hadronic mass
spectrum implies that there is a maximum temperature in a hadronic
gas~\cite{hagedorn}. As we put more energy into the hadron gas, the
energy can go into creating ever more higher mass states instead of increasing
the temperature. Based on this observation, one could wonder, what
existed in the universe when it was hotter than the limiting hadronic
temperature? The answer to that question is provided by the theory of
Quantum Chromodynamics.

\begin{figure}[htb]
  \centering
  \resizebox{0.75\textwidth}{!}{\includegraphics{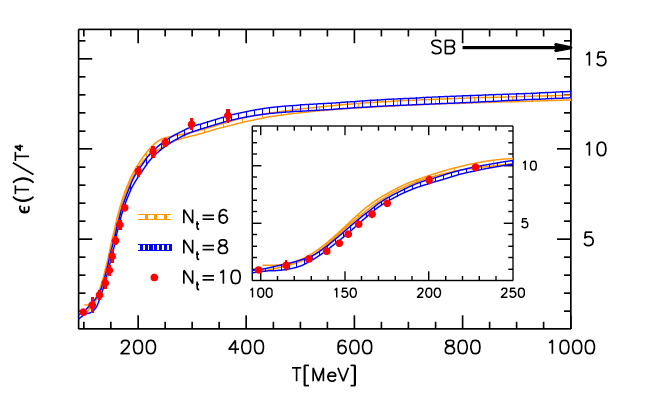}}
  \caption[]{ Lattice calculations of the energy density scaled by
    T$^4$ showing the transition to QGP~\cite{eos}. The energy density of a
    hadron gas would grow to infinity as the temperature approached
    the Hagedorn temperature but instead, the hadronic matter
    dissolves into QGP with a finite number of degrees-of-freedom.}
\label{f2}
\end{figure}

With the discovery that hadrons are made of quarks and gluons
and the development of QCD, the theory that describes the interaction
of quarks and gluons, it becomes clear what happens above Hagedorn's
limiting temperature for hadronic matter. That matter undergoes a
phase transition into a plasma of color charged quarks and gluons
called the quark gluon plasma (QGP)~\cite{Shuryak:1980}. The theory of
QCD has been tested and verified in the high energy, short wavelength,
perturbative limit but the QGP is dominated by non-perturbative
interactions. In this case, we rely on computationally intense Lattice
QCD calculations to study the transition from hadronic matter to
QGP~\cite{eos}. Figure~\ref{f2} shows a Lattice QCD calculation of
energy density scaled by the fourth power of temperature versus the
temperature. This quantity is proportional to the number of degrees of
freedom. As expected, as the temperature is increased, the number of
degrees of freedom begins to rapidly increase. This is the effect
expected from the rise in the number of more massive hadronic states
that become accessible at higher energies. Then, as the energy
continues to grow, the number of degrees of freedom levels off and
nearly saturates. That saturation is a clear demonstration of a
transition into a QGP where the number of degrees of freedom is
determined by the number of quarks and gluons, not by the ever
increasing number of hadronic states. These calculations demonstrate
that QGP is a state of matter predicted by QCD. The task undertaken by
RHIC and other heavy ion experimental facilities~\cite{wp} is to
recreate that state and study its properties.

\section{RHIC Collisions}

The creation of a QGP at RHIC is typically initiated by colliding
together Gold nuclei at $\sqrt{s_{_{NN}}}=200$~GeV, the maximum center
of mass energy acheivable at RHIC for those species. The RHIC
facility is flexible however and can collide a range of nuclei from
single protons up to even heavier elements like Uranium 238 and with
energy ranges from approximately 5 GeV to 200 GeV (or higher for lighter nuclei up to 500
GeV for protons). With this broad flexibility and a suite of detector systems
(the PHOBOS, BRAHMS, STAR and PHENIX detectors~\footnote{the smaller experiments PHOBOS and BRAHMS were decomissioned in 2005}), RHIC is the
ideal facility to study the transition from hadronic
matter to QGP.

\begin{figure}[htb]
  \resizebox{0.59\textwidth}{!}{\includegraphics{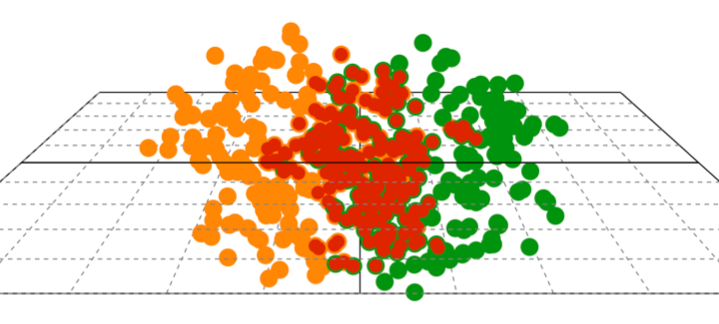}}
  \resizebox{0.39\textwidth}{!}{\includegraphics{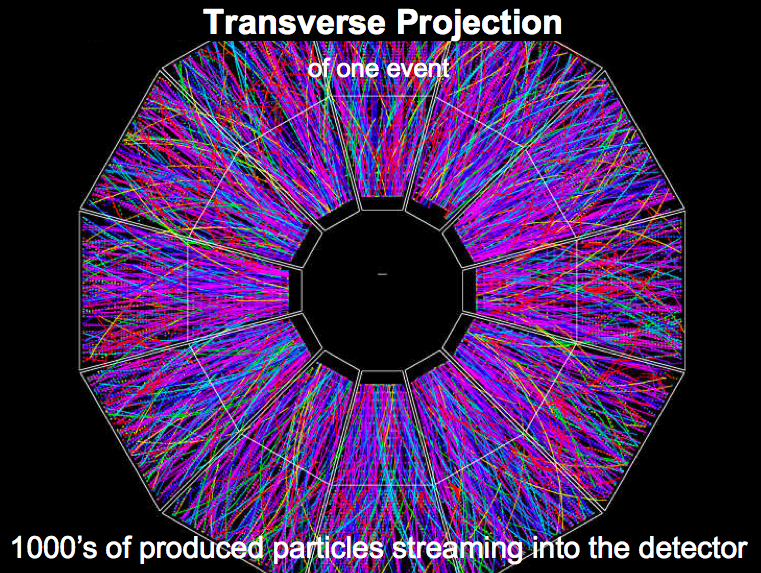}}
  \caption[]{ Transverse view of the initial overlap region and a
    transverse view of the particles streaming from the collision into
    the detector. }
\label{f3}
\end{figure}

Figure~\ref{f3} (left) shows a Monte-Carlo Model of the collision of
two highly Lorentz contracted Gold nuclei~\cite{glauber}. For
collisions at 200 GeV, the Gold nuclei are contracted along their
direction of motion by a factor of 100. The illustration shows the
protons and neutrons distributed within each nucleus --- those that
are expected to participate in the collision are shown in red. On the
right, is an image from the STAR detector~\cite{STAR} showing the
tracks left behind by thousands of charged particles produced in a 200
GeV Au+Au collision. From these tracks, we attempt to infer the
properties of the tiny speck of matter that briefly existed after the
collision. The size of the matter is approximately the size of a Au
nucleus (10$^{-14}$ meters or 10 fm), and it exists for $t \approx 20$
fm/$c$ ($\approx 10^{-22}$ seconds) before the interactions between the
fireball's constituents cease and hadrons freely stream to the
detectors. Since it is impossible to directly observe the matter that
is formed, we look at the abundances, and distributions of various
particles species, and the patterns with which they are emitted to
infer the properties of the matter.

\section{RHIC Highlights}

\subsection{Particle Ratios and Chemical Freeze-out}

\begin{figure}[htb]
  \centering
  \resizebox{0.6\textwidth}{!}{\includegraphics{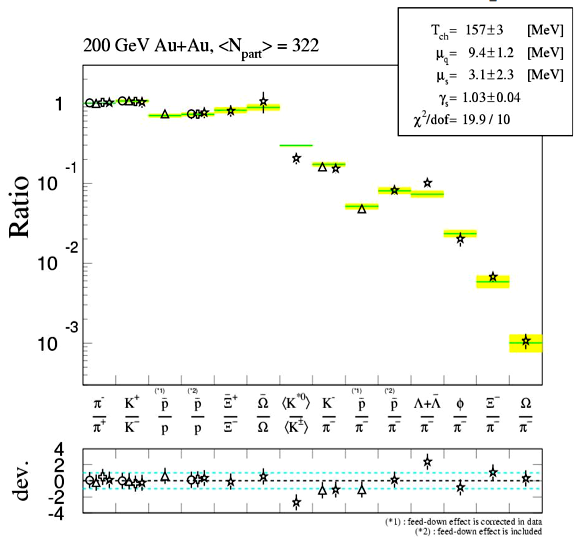}}
  \caption[]{ The ratios of various particle species produced in
    heavy-ion collisions compared to expectations of thermal
    equilibrium~\cite{ratios}. }
\label{f4}
\end{figure}

One of the firsts questions asked is whether regions of the fireball
were locally able to achieve equilibrium before the system blasted
apart. We specify \textit{local} because causality dictates that the
system cannot be completely equilibrated until it has lived for
$c_{s}\tau = R$, where $c_{s}$ is the speed of sound in the medium and
$R$ is the size of the system. Evidence for local thermal
equilibration can be found by investigating the ratios of various
particle species. A thermal model can be constructed to predict these
relative abundances based on simple thermodynamic
arguments~\cite{thermal}. For a thermally equilibrated system, the
relative abundances of particles should only depend on the temperature
of the system, the chemical potentials of the quarks and the masses of
the particles. Figure~\ref{f4} shows a fit of such a model to the
particle ratios measured at RHIC. The fit demonstrates that particle
production from 200 GeV Au+Au collisions is consistent with that
expected from an equilibrated medium, thus satisfying a basic
requirement for the formation of a state of matter.

\subsection{Anisotropy: from Coordinate to Momentum Space}

\begin{figure}[htb]
  \centering
  \resizebox{0.45\textwidth}{!}{\includegraphics{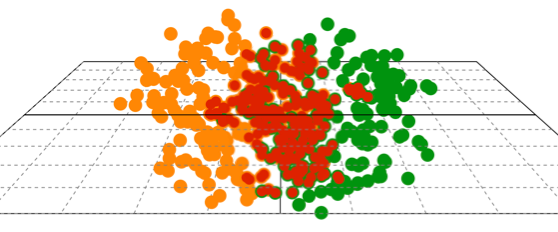}}
  \resizebox{0.45\textwidth}{!}{\includegraphics{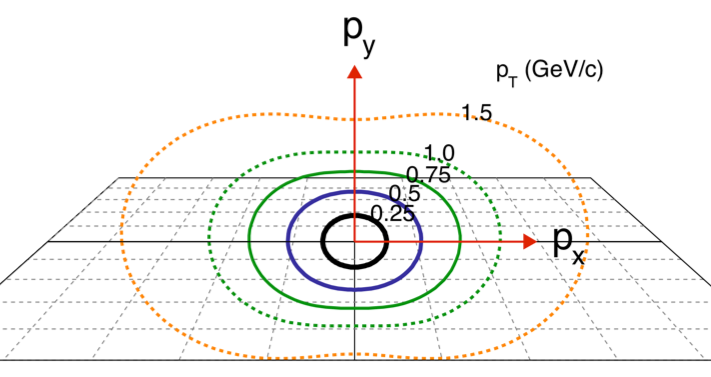}}
  \caption[]{ Conversion from coordinate space into momentum space. }
\label{f5}
\end{figure}

If interactions in the fireball are strong enough to create a locally
thermalized medium, then we may expect the interactions to influence
the patterns with which particles are emitted. Typically, the region
where the two nuclei overlap is not isotropic; for non-central
collisions, it possesses an elliptic shape. If particles free stream
from that region, the elliptic shape will not be transferred into
momentum space, and the particles will be emitted isotropically.
Interactions, on the other hand, will lead to anisotropies in the
momentum space distributions of particles. In particular, if we
consider the pressure gradients in the system, we see
that the largest pressure gradient will be along the short axis of the
elliptic overlap region (see figure~\ref{f5}). That is the axis where
the density goes most abruptly from the maximum at the center to the
minimum at the edge. If pressure gradients are acting on the emitted
particles, that anisotropy in the pressure gradient will lead to an
elliptic anisotropy in the momentum space distribution of emitted
particles (as shown in the right panel of
figure~\ref{f5})~\cite{v2papers}. That elliptic anisotropy is
quantified by $v_2$, the second term in a Fourier Series expansion of
the angular dependence of particle emission in the transverse plane,
where the azimuthal angles are taken relative to the axis connecting
the two colliding nuclei. The right panel of figure~\ref{f5}
illustrates how anisotropic the distributions are for particles
emitted with different transverse momenta. In semi-central Au+Au
collisions at 200 GeV, there are nearly twice as many 1.5 GeV
particles emitted in the direction of the short axis of the overlap
region as there are in the direction of the long axis: note this ratio
will go as $\frac{dN_{\mathrm{short}}}{dN_{\mathrm{long}}}=\frac{1+2v_2}{1-2v_2} \approx
1+4v_2$ so that $v_2=0.25$ corresponds to
$\frac{dN_{\mathrm{short}}}{dN_{\mathrm{long}}}\approx2$. This is a remarkably large
deviation from expectations of free-streaming that indicates the
fireball maintains a memory it's original geometry.

\subsection{Elliptic Flow and Quark Number Scaling}

\begin{figure}[htb]
  \centering
  \resizebox{0.45\textwidth}{!}{\includegraphics{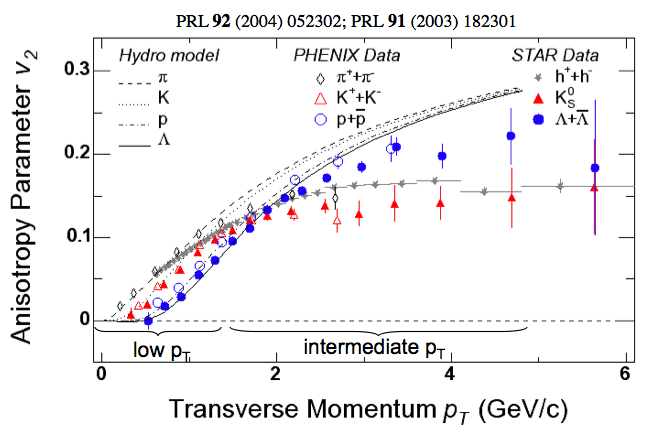}}
  \resizebox{0.45\textwidth}{!}{\includegraphics{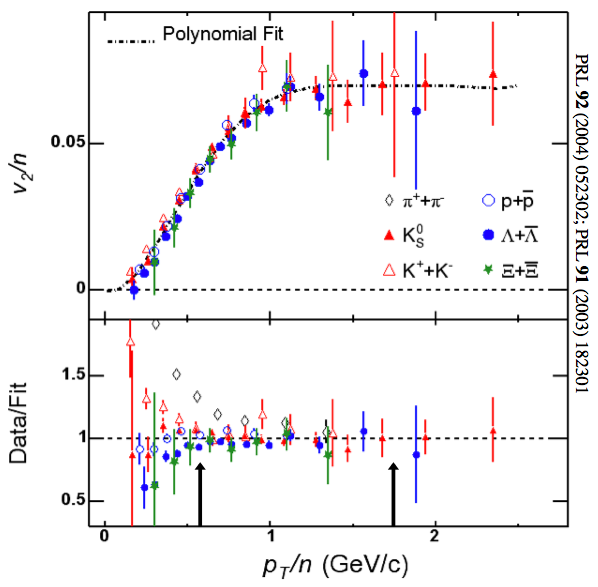}}
  \caption[]{ The momentum-space anisotropy and the same scaled by the
    number of quarks in each hadron. }
\label{f6}
\end{figure}

Measurements of $v_2$ for different particle species at different
transverse momenta are shown in figure~\ref{f6}~\cite{v2papers}. In
the region below $p_T \sim 2$~GeV/$c$, $v_2$ follows mass ordering with
heavier particles having smaller $v_2$ at a given $p_T$. Above this
range, the mass ordering is broken and the heavier baryons take on
larger $v_2$ values. A hydrodynamic model for $v_2(p_T)$ is also shown
which describes the $v_2$ in the lower $p_T$ region well~\cite{hydro}.
The hydrodynamic model assumes local thermal equilibrium and zero
mean-free-path for interactions. The mass ordering in the model and
data are a feature expected for particle emission from a boosted
source. In the case that particles move with a collective velocity,
more massive particles will receive a larger $p_T$ kick. As the
particles are shifted to higher $p_T$, the lower momentum regions
become depopulated with a larger reduction in the direction with the
largest boost (in-plane). This reduction reduces $v_2$ at a given
$p_T$, with the reduction largest for more massive particles, as
observed in the data. This is clear evidence for a fireball expanding
under the influence of anisotropic pressure gradients. The successful
description of the $v_2$ data in the momentum range where 99\% of the
particles are produced with a ideal hydrodynamic models led physicists
at RHIC to dub the matter a perfect liquid~\cite{perfect}. In fact
preliminary estimates of the viscosity of the QGP created at RHIC
indicate that it is lower than any known substance and consistent with
the lowest value possible~\cite{bound}.

At higher $p_T$, $v_2$ no longer rises with $p_T$ and the mass
ordering is broken. Above $p_T$ near 2~GeV/$c$ the more massive baryons
exhibit a larger $v_2$ than the mesons. While the pion and kaon $v_2$
reach a similar maximum of $v_2 = 0.14$ at $p_T$ near 2.5~GeV/$c$, the baryon $v_2$ continues to rise until it reaches a
maximum of $v_2 = 0.20$ at $p_T$ near 4.0~GeV/$c$. For still
larger $p_T$, the $v_2$ values exhibit a gradual decline until $v_2$
for all particles is consistent with $v_2 = 0.10$ at $p_T$ near 7~GeV/$c$ (not shown in this figure). At these higher $p_T$
values one expects that the dominant process giving rise to $v_2$ is
jet-quenching~\cite{Wang:1991xy} where hadron suppression is larger
along the long axis of the overlap region than along the short
axis~\cite{Wang:2000fq,Gyulassy:2001kr,Gyulassy:2000gk}. Jet quenching
occurs as fast moving partons radiate energy in the expanding medium.
This leads to a dramatic suppression of high momentum particles which
will be discussed later. For very large energy loss, the value of
$v_2$ should be dominated by the geometry of the collision region.

\begin{figure}[htb]
  \centering
  \resizebox{0.7\textwidth}{!}{\includegraphics{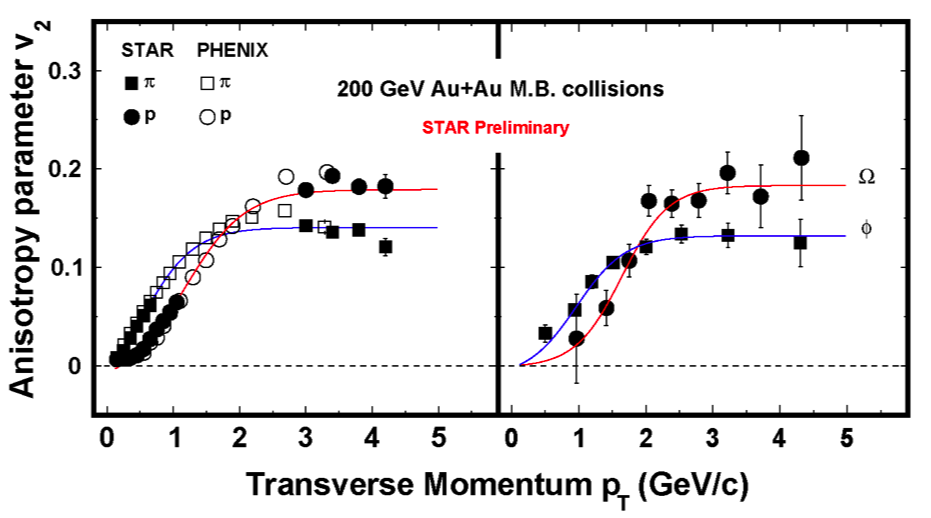}}
  \caption[]{ Elliptic flow of strange and multiply strange hadrons
    showing that the quark number scaling also holds for particles
    with larger masses and that the $\phi$-meson groups with other
    particles having the same number of constituent quarks, rather
    than with particles having similar masses like protons. }
\label{f7}
\end{figure}

At an intermediate $p_T$ range, it was noted that mesons and baryons
spread out into two bands with the generally heavier baryons
developing a larger $v_2$, contrary to the behavior in the lower
momentum region. It was also noticed that if the $v_2$ and $p_T$ of
the hadron were scaled by the number of constituent quarks in the
hadron, the $v_2/n$ vs $p_T/n$ of various species all lined up with
one another~\cite{Fries:2008hs}. This scaling is shown in the right
panel of figure~\ref{f6}. Figure~\ref{f7} shows that this grouping of
mesons and baryons extends to the heavier, multiply strange hadrons,
the $\phi$-meson and $\Omega$-baryon. The data in figure~\ref{f7} is
important because it shows that although the $\phi$-meson is as heavy
as the proton, it behaves like other mesons. Indeed, the number of
constituent quarks seems to be more relevant than the mass of the
hadron.

\subsection{The Baryon Enhancement}

The observation of the quark-number dependence of $v_2$ at
intermediate $p_T$ and the constituent quark number scaling led to
speculation that hadron formation was happening via the coalescence of
dressed quarks at the hadronization phase boundary, leading to an
amplification of $v_2$ with baryons getting amplified by a factor of 3
while mesons were amplified by a factor of
two~\cite{Fries:2008hs,Voloshin:2002wa,Lin:2001zk,Molnar:2003ff,Fries:2008hs,Hwa:2003bn,Fries:2003kq,Greco:2003mm,Greco:2003xt,Fries:2003vb,Pratt:2004zq,Ravagli:2008rt}.
This picture was subsequently strengthened by the observation that a
similar quark-number dependence arises in
$R_{CP}$~\cite{Adams:2003am,Adler:2003kg}: the ratio of the single
particle spectra in central collisions to that in peripheral
collisions. At intermediate $p_T$ the $R_{CP}$ values for various
particle species are also grouped by the number of constituent quarks,
with baryons having a larger $R_{CP}$. The larger $R_{CP}$ for baryons
signifies that baryon production increases with collision centrality
faster than meson production; an observation consistent with the
speculation that hadrons from Au+Au collisions are formed by
coalescence such that baryon production becomes easier as the density
of the system increases. The more general and less model dependent
statement is that the baryon versus meson dependence arises from high
density and therefore most likely from multi-quark or gluon effects or
sometimes called "higher twist" effects.

\begin{figure}[htb]
 \centering
 \resizebox{0.9\textwidth}{!}{\includegraphics{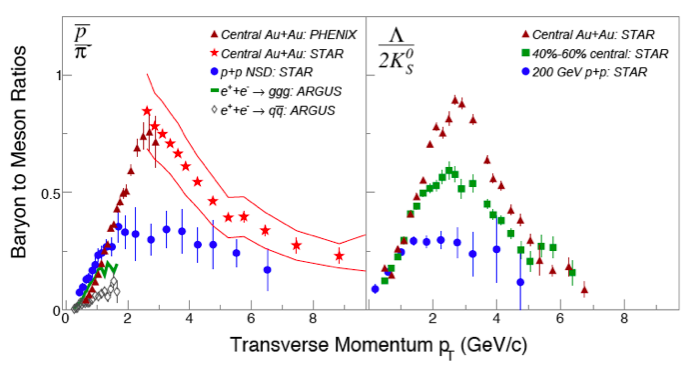}}
 \caption[]{ Baryon to meson ratios in 200 GeV Au+Au and p+p
   collisions. The data shows that the denser Au+Au collisions
   produce more baryons relative to mesons than the less central
   Au+Au collisions or p+p collisions. }
\label{f8}
\end{figure}

 The combination of large baryon $v_2$ and large baryon $R_{CP}$ also
immediately eliminates a class of explanations attempting to describe
one or the other observation: \textit{e.g.} originally it was
speculated that the larger $R_{CP}$ for baryons might be related to a
smaller jet-quenching for jets that fragment to baryons than for jets
that fragment to mesons. This explanation would lead to a smaller
baryon $v_2$ and is therefore ruled out by the larger $v_2$ for
baryons. The same can be said for color transparency
models~\cite{Brodsky:2008qp} which would account for the larger baryon
$R_{CP}$ in this $p_T$ region but would predict a smaller baryon
$v_2$. Color transparency may still be relevant to the particle type
dependencies at $p_T>5$ where $R_{CP}$ for protons is slightly larger
than $R_{CP}$ for pions~\cite{Abelev:2007ra} and the $v_2$
measurements are not yet precise enough to conclude whether the baryon
$v_2$ is also smaller than the meson $v_2$. This is a topic that needs
to be studied further. Figure~\ref{f8} shows baryon to meson ratios
for p+p collisions, mid-central Au+Au collisions, and central Au+Au
collisions. At intermediate $p_T$ values, the number of baryons
produced is similar to the number of mesons. $v_2$ measurements also
demonstrate that baryons are more abundantly produce along the short
axis of the overlap region rather than perpendicular to it. The two
observations are directly related to each other.

\subsection{Correlations and Fluctuations}

\begin{figure}[htb]
\centering
\resizebox{0.9\textwidth}{!}{\includegraphics{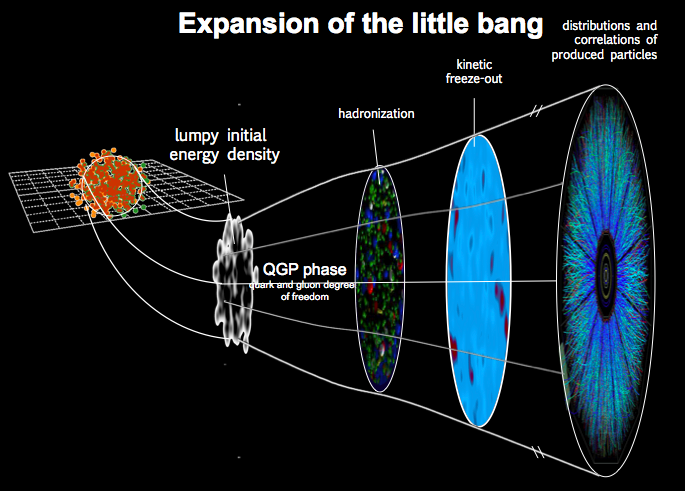}}
\caption[]{ A sketch of the expansion of a heavy ion collision. }
\label{f9}
\end{figure}

The data presented above along with many other observations have led
to a picture of heavy-ion collisions summarized in the diagram in
figure~\ref{f9}. Energy is deposited in the overlap region of the two
colliding nuclei. That matter reaches local thermal equilibrium in a
surprisingly short time (about 1 fm/c) and the quark-gluon-plasma
phase of the expansion begins. The QGP expands in the longitudinal and
transverse directions developing a hubble-type expansion with
transverse velocity $v_r \propto r$. The system quickly cools with
cooler regions passing through the phase transition to hadronic
matter. Eventually, all of the QGP has converted to hadronic matter.
Later still, inelastic interactions cease and the particle ratios are
fixed. This is called chemical freeze-out. Finally the elastic
interactions cease and the momentum of the particles cease to change.
This is called kinetic freeze-out. After this, the particles stream
freely to the detector.


\begin{figure}[htb]
\centering
\resizebox{0.8\textwidth}{!}{\includegraphics{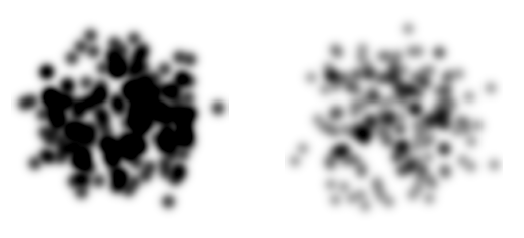}}
\caption[]{ Gluon density in the transverse plane when the nucleus is
 probed at different $x$ values by a 0.2 fm quark-antiquark dipole in
 the IPsat CGC model. }
\label{lumpy}
\end{figure}

 It is apparent from the $v_2$ measurements that during the evolution
sketched in figure~\ref{f9}, the fireball converts features of the
initial overlap geometry into momentum space. The initial overlap
region, however, is neither a perfect sphere nor a perfect ellipse. It
is expected to contain regions of high and low energy density. As an
example, figure~\ref{lumpy} shows the gluon
density in the transverse plane probed by a 0.2 fm quark-antiquark
dipole for two different $x$ values in the IPsat Mean-Field
theoretical model of QCD~\cite{Kowalski:2007rw} ($x =
2p_T/\sqrt{s_{_{NN}}}$ is $10^{-5}$ in the left panel and $10^{-3}$ in
the right panel). A lumpiness is immediately apparent because the
gluons are confined in the transverse plane to regions near the
valence charges in the nucleons within the nucleus. Although this
picture leads to longitudinally extended flux-tubes due to the overlap
of the wave-functions in the longitudinal direction, in the transverse
direction, there is a great deal of inhomogeneity. One can ask if the
fireball transfers those features into the final momentum space
distributions. If so, then they should show up as correlations between
produced particles that are extended in the longitudinal direction but
narrow in the azimuthal direction.

\begin{figure}[htb]
  \centering
  \resizebox{0.99\textwidth}{!}{\includegraphics{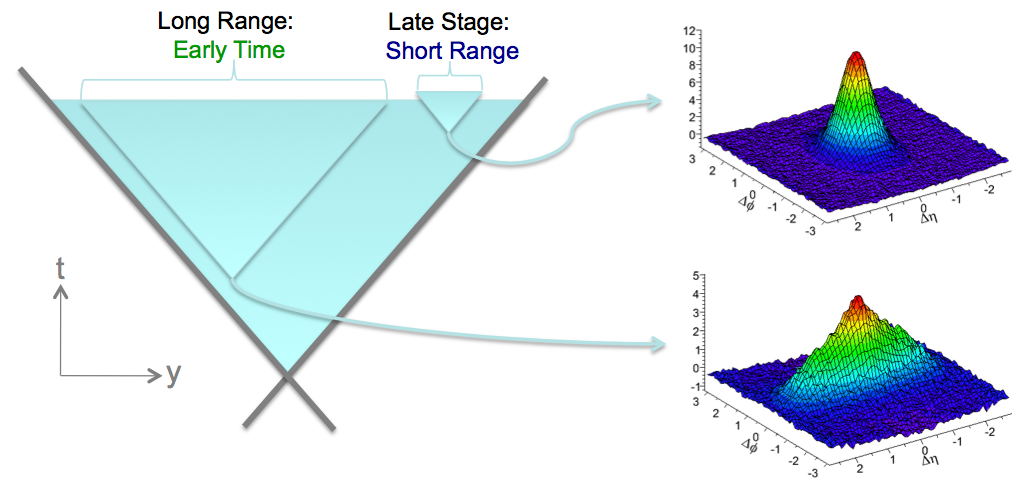}}
  \caption[]{ The relationship of long and short range correlations to
    early and late times~\cite{Mocsy:2009qd}. }
\label{f10}
\end{figure}

There are many conjectured processes that can lead to correlations
between the produced particles; resonance decays, fragmentation of
jets, bubble nucleation or spinodal decomposition at a phase boundary~\cite{csernai,randrup,Mocsy:2009qd},
and quantum interference are some of the sources of correlations
frequently considered. Of all these correlations, only correlations
from the very early stages of the evolution can spread out very far in
the longitudinal direction. The reason for this is illustrated in
figure~\ref{f10}. The left of the figure shows the space time picture
of a heavy ion collision. The trajectories of the incoming nuclei and
the outgoing nucleons moving near the speed of light are shown as
thick grey lines. The horizontal axis shows the longitudinal rapidity
y and the vertical axis shows time. Causality dictates that
correlations at later times, cannot spread out far in
rapidity~\cite{Dumitru:2008wn}. Only correlations from early in the
evolution can stretch far in rapidity. One way to look for
correlations left over form the lumpiness in the initial overlap
region, is to look for correlations that are extended in the
longitudinal direction.

\begin{figure}[htb]
  \centering
  \resizebox{0.99\textwidth}{!}{\includegraphics{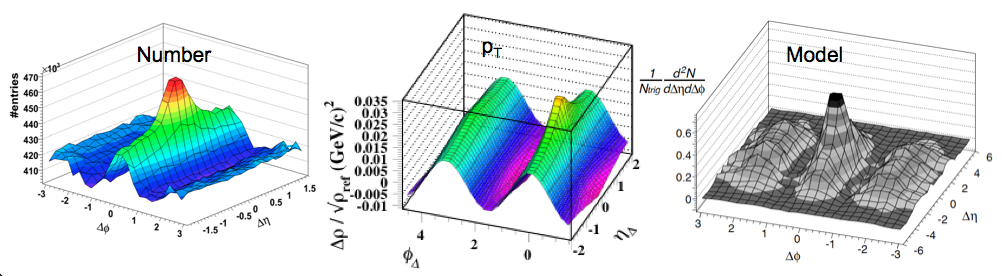}}
  \caption[]{ Measured and calculated long range correlations. }
\label{f11}
\end{figure}

The left panel of figure~\ref{f11} shows measurements of two particle
correlations where one particle is required to have transverse momentum $p_T$ above 4
and the other above 2 GeV/$c$~\cite{ridgedata}. These selections are
usually made to search for correlations arising from jets but they
exhibit decidedly non-jet-like structures. A jet will lead to
correlations that are narrow in the azimuthal and longitudinal
direction. Such a correlation is apparent in the peak at $\Delta\eta$
and $\Delta\phi$ near zero where $\eta$ is the longitudinal
pseudo-rapidity variable. In addition though, one notices a
correlation which extends in $\Delta\eta$ across the entire
acceptance of the detector. This ridge-like correlation structure has
the characteristics one would expect from lumpiness in the initial
overlap zone.

The middle panel of figure~\ref{f11} shows two particle $p_Tp_T$
correlations between all observed hadrons (with no $p_T$
selection)~\cite{Adams:2005aw}. This data shows the correlations
expected from $v_2$ (a $\cos(2\Delta\phi)$ modulation) with an
additional peak centered at $\Delta\eta$ and $\Delta\phi=0$ which is
also broadened longitudinally. The model calculations shown in the
right panel are from a hydrodynamic model with a lumpy initial overlap
density as the initial conditions~\cite{Takahashi:2009na}. The
granular nature of the initial conditions clearly show up after the
hydrodynamic expansion as correlations similar to those seen in the
data. This comparison therefore seems to support the picture
of heavy-ion collisions described above, where a hubble-like expansion
converts the lumpiness in the initial energy density into correlations
that are observed in momentum space.

\subsubsection{From $v_{2}$ to $v_n$: the Power Spectrum of Heavy Ion
  Collisions}

For several decades,emphasis has been placed on measurements of
the second harmonic of the particle distributions in the azimuthal
direction $v_2$. This is natural since the dominant shape of the
overlap region in non-central collisions is an ellipse. But
understanding the finer detail in the overlap region has become
increasingly important: both as a way to resolve previous
uncertainties and as a way to learn something new about the
collisions. By now it is clear that rather than only studying $v_2$ we
should be studying the spectrum $v_n$ vs $n$, particularly in very
central collisions where the elliptic shape of the overlap area
subsides so that the second harmonic doesn't overwhelm the others~\cite{qmvnrefs}.
This spectrum is analogous to the power spectrum that is extracted
from temperature-temperature correlations measured in the cosmic
microwave background radiation (CMB). This analogy was first pointed
out in 2007 in a paper by Mishra et. al. where the authors proposed
the analysis of $v_n^2$ for all values of $n$ as an analogous
measurement to the Power Spectrum extracted from the
CMB~\cite{Mishra:2007tw}. They argued that density inhomogeneities in
the initial state would lead to non-zero $v_n^2$ values for higher
harmonics including odd terms like $v_3$ that previously had been
assumed to be zero by symmetry. It was subsequently pointed out that
information on $v_n^2$ was to a large extent already contained within
existing two-particle correlations data~\cite{Sorensen:2008dm}, and
that $v_n$ and $v_n$ fluctuations would provide a natural explanation
for the novel features seen in those correlations, such as the ridge
like~\cite{softridge,ridgedata} structure as well as double hump structure on
the away side that had been attributed to a mach-cone~\cite{awayside}.

That the ridge could be related to flux-tube like structures in the
initial state was already argued by Voloshin in 2006~\cite{radflow}.
This was bolstered by hydrodynamic calculations carried out within the
NEXSPHERIO model shown in figure~\ref{f11} (right) which showed that
fluctuations in the initial conditions lead to a near-side ridge
correlation and a mach-cone like structure on the
away-side~\cite{Takahashi:2009na}. In 2010, Alver and Roland used a
generalization of participant eccentricity
($\varepsilon_{n,\mathrm{part}}$) to arbitrary values of $n$ (as had
been done previously in Ref.~\cite{broniowski}) to show that within
the AMPT model, the final momentum space anisotropy for $v_3$ is
proportional to the initial $\varepsilon_{3,\mathrm{part}}$~\cite{AR}.
This explained the previous observation that the AMPT model produced
correlations similar to those seen in the data (albeit with smaller
amplitudes)~\cite{Ma:2006fm}. This history of studies motivates the
analysis of two-particle correlations in terms of $v_n$ harmonics.

\begin{figure}[htb]
\centering
\resizebox{0.95\textwidth}{!}{\includegraphics{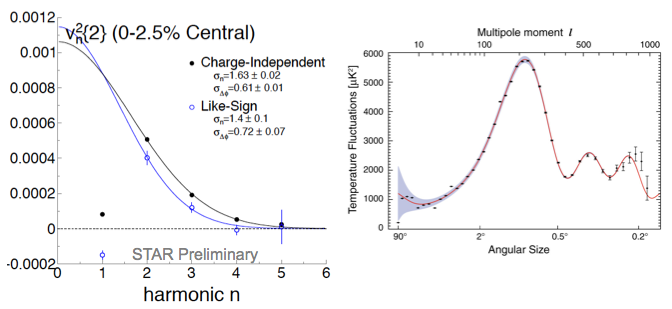}}
\caption[]{ Left: The Fourier Transform of two-particle correlations
 in 0-2.5\% central Au+Au collisions; a measurement analogous to the
 power spectrum measured from the Cosmic Microwave Background
 Radiation (shown on the right). } \label{fps}
\end{figure}
 
The left panel of figure~\ref{fps} shows $v_n\{2\}^2$ versus $n$ for
0-2.5\% central collisions. The $v_n\{2\}^2$ results are taken from
the Q-cumulant analysis which integrates over either all pairs of
particles within $|\eta|<1$ and $0.15 < p_T < 2.0$ GeV/$c$ independent
of charge or for only like-sign pairs. The two spectra show similar
behavior with the $n=1$ components suppressed and the remainder of the
spectra falling off quickly. Excluding $n=1$, The drop with $n$ is
well described by a Gaussian. The Gaussian width for $v_n\{2\}^2$ vs.
$n$ from all pairs independent of charge is $\sigma_{n}=1.63\pm 0.02$
as expected from the observed $\Delta\phi$ width of the near-side peak
in the two-particle correlations~\cite{softridge}. If $v_n\{2\}^2$ is
sensitive to the initial eccentricity, $v_1\{2\}^2$ should be
suppressed because $\varepsilon_{1,\mathrm{part}}^2\approx 0$ (due to
the constraint that the eccentricity is defined relative to the
center-of-mass of the participating nucleons~\cite{rise}). Higher
harmonics are expected to be washed out by viscous and other effects
that smear out fine structure~\cite{sound}. The shape of $v_n\{2\}^2$ is therefore
a sensitive probe of viscous effects in heavy ion collisions.

The $v_n\{2\}^2$ spectrum from heavy-ion collisions can be compared
and contrasted with the Power Spectrum from the CMB. One obvious
difference is that viscous effects are not important in the early
universe, owing to the large size of the universe compared to the
viscous length scales, while in heavy-ion collisions, length scales
like the mean-free-path are commensurate with the size of the system.
As such, the Power Spectrum from the CMB extends to much higher
harmonics whereas the $v_n\{2\}^2$ spectrum in heavy ion collisions
drops off very quickly with harmonic $n$. The CMB power spectrum also
exhibits a local maximum near Multi-pole moment $l=200$ followed by
several oscillations. These features arise because fluctuations at
length scales larger than the acoustic horizon are suppressed in the
CMB. No such feature is seen in the $v_n\{2\}^2$ spectrum for the
kinematic selections shown here. Although Mishra et. al. proposed the
measurement of $v_n$ vs. $n$ in order to observe the effect of
super-horizon fluctuations, it was subsequently realized that higher
harmonics of $v_n$ could be washed out by viscous effects and that the
shape of $v_n$ vs. $n$ is more related to
these effects~\cite{sound}.

An acoustic horizon suppresses lower harmonics while viscous-type
effects suppress higher harmonics. Viscous effects appear to dominate
the shape of the $v_n$ spectrum for low momentum particles but the
effect of the horizon may have already been observed at higher $p_T$
where $v_n\{2\}^2$ shows a local maximum at $n=3$. This result may be
explained by a hubble expansion and the suppression of lower harmonics
which are super horizon. If the expansion velocity increases linearly
with the radius as with a hubble expansion, higher $p_T$ particles
tend to be emitted from the edge of the system. Hydrodynamic
calculations show that this type of radial expansion should be
established in a time of about 5 fm/$c$~\cite{kolb}. The edge of the
collision region is also where horizon effects should be most
prominent. The harmonic number below which the acoustic horizon will
suppress the power spectrum (corresponding to one fourth of a
wavelength fitting inside the horizon) goes as $n\approx2\pi R/(4H)$
where $R$ is the radius at which the particles were emitted and $H$ is
the horizon. If the average of the emission radius for all particles
is 8 fm (consistent with blast-wave model fits to data~\cite{lisa}),
and the acoustic horizon grows to approximately 7 fm (as estimated
from lattice data and hydro~\cite{sound}), then the peak in the $v_n$
spectrum should be near $n=8\pi/14=1.8$. A peak in the $v_n$ spectrum
at $n=1.8$ will not be observable because $n=1$ is dominated by
momentum conservation so that $v_1$ must be zero when averaged over
all particles. We therefore cannot observe a suppression below $n=2$
due to the horizon since $n=1$ is already suppressed and we can only
evaluate the spectrum in discrete steps.

If the fireball develops a hubble expansion, the edge of the system
will be boosted with a larger velocity and tend to emit particles with
higher $p_T$. When we select high $p_T$ particles, their emission
radius will therefore tend to be larger. In a blast wave model the
average emission radius of particles near $p_T=2$~GeV/$c$ is closer to
12 fm. For this subset of particles, the peak in the spectrum should
be near $n\approx2.7$. This is consistent with the observation of a
local maximum in the $v_n$ spectrum near $n=3$. This
back-of-the-envelope calculation should be verified in a full
simulation of heavy ion collisions. I've also neglected
the possibility of a smaller acoustic horizon on the edge of the
system than in the center but the argument at least appears plausible.
Given that no other explanation of the local maximum at $n=3$
exists, if confirmed in a more rigorous calculation, this will provide
powerful evidence for the existence of a hubble-like expansion in the
evolution, thus adding yet another convincing piece of evidence in
favor of the picture of heavy ion collisions described above.

\begin{figure}[htb]
  \centering
  \resizebox{0.95\textwidth}{!}{\includegraphics{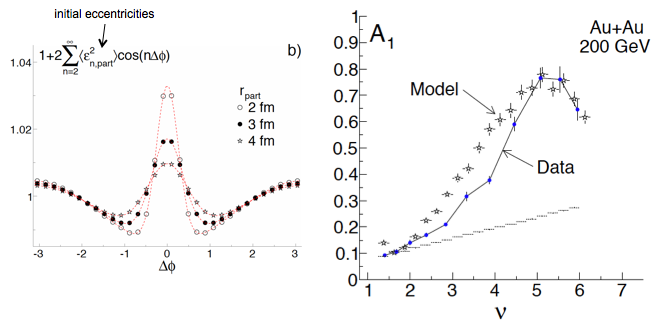}}
\caption[]{ Left: The azimuthal shape of correlations expected from
  eccentricity fluctuations caused by the lumpiness of the initial
  matter density. Right: The amplitude of the Gaussian peak at
  $\Delta\phi=0$ compared to the data. Both the data and the model
  show a rise and then a fall in the most central collisions.}
\label{frf1}
\end{figure}

Since there are several possible sources for the observed
correlations, before interpreting the $v_n\{2\}^2$ spectrum, one would
like to be absolutely sure that the correlations are indeed dominated
by the initial density inhomogeneities. Towards this goal, we can
compare specific characteristics of the observed correlation functions
to expectations for correlations from the initial density
inhomogeneities. Data shows that the amplitude of the near-side ridge
correlation rises rapidly with collision centrality~\cite{softridge}
before reaching a maximum and falling off in the most central
collisions. This rise and fall has been observed at center of mass
energies of 62.4 GeV, 200 GeV, and 2.76 TeV~\cite{lhcridge}. The drop
is often ignored but it shouldn't be; it provides a clear
indication that the correlations arise from the initial state
lumpiness. Reference~\cite{rise} provides the explanation for the rise and
fall of the ridge amplitude by relating it to density inhomogeneities
in the initial overlap region.

By generalizing the eccentricity to any arbitrary value of
$n$~\cite{broniowski,AR}, one can characterize the initial density
distributions via the eccentricity of the participants in the
collision. The participant eccentricity $\varepsilon_{n,part}^2$
represents a harmonic decomposition of the azimuthal dependence of the
initial overlap density. Smearing out the positions of the
participants over some region ($r_{\mathrm{part}}$) washes out higher harmonics
such that the shape of $\varepsilon_{n,\mathrm{part}}^2$ vs $n$ is well
described by a Gaussian centered at $n=0$ (except that $n=1$ will be
zero since the eccentricity is naturally calculated in the
center-of-mass frame). If $v_n^2 \propto \varepsilon_{n,\mathrm{part}}^2$ then
two-particle correlations due to eccentricity fluctuations should have
the form shown in figure~\ref{frf1} (left): the two-particle
correlations vs $\Delta\phi$ are just related to
$\varepsilon_{n,\mathrm{part}}^2$ by a Fourier transform. Based on this
realization, one can make a rough estimate of the centrality
dependence of the near-side ridge due to the eccentricity fluctuations
caused by the lumpiness of the initial density. The right panel of
figure~\ref{frf1} shows that estimate compared to data. Although the
agreement between model and data is not perfect, the model correctly
describes the rise and fall of the ridge amplitude. The fact that the
model provides an explanation for the rise and fall, suggests that the
correlation data is dominated by eccentricity fluctuations. It is
therefore worth delving into what causes the rise and fall in the
eccentricity based model.

\begin{figure}[htb]
\centering
\resizebox{0.85\textwidth}{!}{\includegraphics{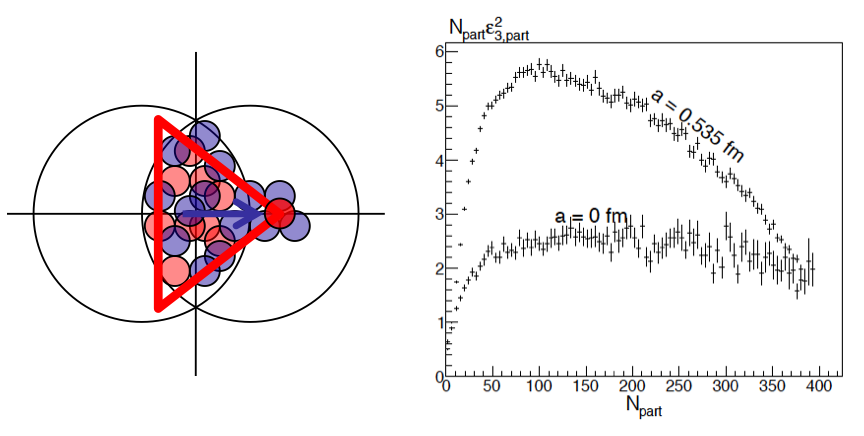}}
\caption[]{ Left: The rise and fall of the ridge is driven by
  fluctuations as illustrated in this figure. Right: Monte Carlo
  Glauber calculations of the mean of the square of the third harmonic
  participant eccentricity scaled by the number of participants
  $N_{\mathrm{part}}\varepsilon_{3,\mathrm{part}}^2$ as a function of
  $N_{\mathrm{part}}$ for Au+Au collisions. Two cases are shown: 1)
  the surface of the Au nucleus is treated as a hard shell so that
  nucleons do not fluctuate beyond the nominal radius of the nucleus
  (Woods-Saxon diffuseness parameter $a = 0$ fm) and 2) the edge of
  the Au nucleus is treated as diffuse so that that nucleons can
  fluctuate beyond the nominal Au radius (Woods-Saxon diffuseness
  parameter $a=0.535$ fm is taken from electron scattering data). }
\label{frf2}
\end{figure}

In the eccentricity based model, the rise and fall of the ridge is
driven by fluctuations as illustrated in figure~\ref{frf2} (left). A
nucleon near the edge of one nucleus impinges on many nucleons from
the other nucleus. This induces participant eccentricity fluctuations
for many harmonics leading to a near-side Gaussian peak in
two-particle correlations. For very central collisions, a nucleon
fluctuating beyond the edge of a nucleus will not likely encounter a
nucleon from the other nucleus. In that case, this particular source
of fluctuations will subside. The subsidence of that source of
fluctuations is seen in the two-particle correlations as the drop-off
of the ridge amplitude $A_1$ in the most central collisions.

This detailed explanation for the rise and fall of the ridge
demonstrates that the ridge in the two-particle correlations data is
indeed caused by fluctuations in the initial density of the
collisions. In the right panel of figure~\ref{frf2}, Monte
Carlo Glauber calculations of the mean of the square of the third
harmonic participant eccentricity scaled by the number of participants
$N_{\mathrm{part}}\varepsilon_{3,\mathrm{part}}^2$ are shown as a function of
$N_{\mathrm{part}}$ for Au+Au collisions. Two cases are given: In one
case, the surface of the Au nucleus is treated as a hard cut-off so
that nucleons do not fluctuate beyond the nominal radius of the Au
nucleus (Woods-Saxon diffuseness parameter $a = 0$ fm) and in the
other case, the edge of the Au nucleus is treated as diffuse so that
nucleons can fluctuate beyond the nominal Au radius (Woods-Saxon
diffuseness parameter $a=0.535$ fm taken from electron scattering
data~\cite{devries}). The $a=0$ fm calculation artificially suppresses the
fluctuations responsible for the rise and fall of the ridge. This case
would lead to a ridge that increases linearly with centrality. The
physically meaningful calculation with $a=0.535$ fm leads to a ridge
that first rises abruptly and then falls in the most central
collisions. The prediction for the amplitude of the ridge shown in
figure~\ref{frf1} is based on the centrality dependence of the ridge
amplitude $A_1$ following
$\rho_0N_{\mathrm{part}}\varepsilon_{n,\mathrm{part}}^2$ where
$\rho_0$ is the particle density. This estimate correctly predicted
the centrality dependence of the ridge in Pb+Pb collisions at 2.76
TeV prior to those measurements being made. There seems to be little doubt that the observed two-particle
correlations are dominated by lumpiness in the initial density
distributions.

\begin{figure}[htb]
  \centering
  \resizebox{0.99\textwidth}{!}{\includegraphics{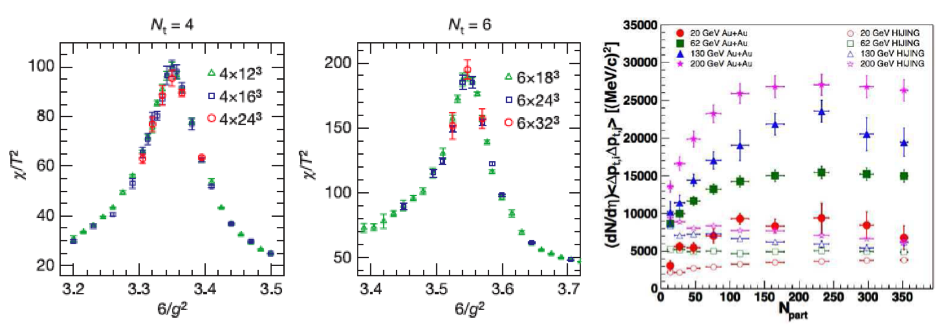}}
  \caption[]{ The size dependence of susceptibilities in lattice
    calculations~\cite{Aoki:2006we} and the measured pt fluctuations
    in data~\cite{Adams:2005ka}. }
\label{f12}
\end{figure}

Besides the dominant correlations that have been demonstrated to arise
from the lumpiness of the initial density distributions, correlations
and fluctuations can also develop at the phase boundary between QGP
and hadronic matter. If the phase transition is second or first order
we would expect the correlations to be larger for larger systems.
These correlations and fluctuations are related to the
susceptibilities in the system. Lattice QCD calculations show that the
magnitude of the peak in the susceptibilities at the phase boundary
does not change when the size of the lattice changes. This indicates
that the phase transition in QCD at zero baryon density is a smooth
cross-over~\cite{Aoki:2006we}. In data, as discussed in the previous
paragraphs, the measured correlations and fluctuations seem to be
dominated by the initial conditions rather than by correlations and
fluctuations from the phase boundary. The right panel of
figure~\ref{f12} shows $p_Tp_T$ fluctuations for several different
energies vs an estimate of the number of nucleons participating in the
collision~\cite{Adams:2005ka}. Above about 100 participants, the
fluctuations are independent of the system size. This is consistent
with the smooth crossover from QGP to hadron gas as expected from
Lattice QCD calculations.

\subsection{The RHIC Beam Energy Scan}

\begin{figure}[htb]
 \centering
 \resizebox{0.7\textwidth}{!}{\includegraphics{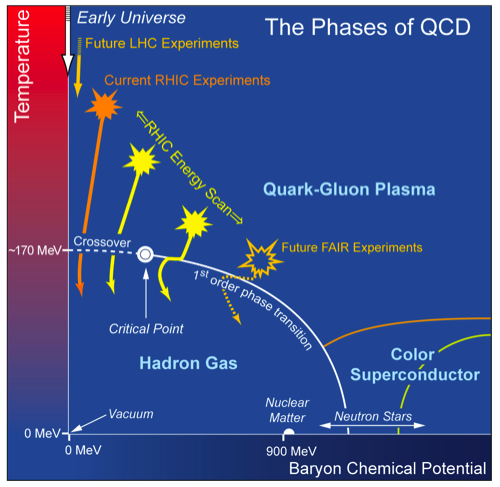}}
 \caption[]{ The phase diagram of nuclear matter. }
\label{f13}
\end{figure}
 
Although Lattice calculations show that the QGP to hadron phase
transition is a smooth crossover at zero baryon density, it is thought
to be a first-order phase transition at higher baryon
densities~\cite{criticalpoint}. Lattice calculations suffer from the
sign problem at non-zero baryon chemical potential so it is not easy
to infer from QCD where the transition from a smooth crossover to a
first order phase transition occurs. The point where the smooth
crossover and first order phase transition meet is called a critical
point and represents an important landmark on the phase diagram of
nuclear matter. Figure~\ref{f13} shows a schematic diagram of the
phases of QCD. Full energy heavy-ion collisions at RHIC and those that
at the LHC, are nearly net baryon free and probe along the left axis
of the diagram. The flexibility of RHIC makes it possible to probe
deeply into the non-zero baryon chemical potential axis. By lowering
the energy of the colliding beams, one initiates a system with larger
net baryon density. By scanning down to lower and lower energies, the
RHIC experiments can search for evidence of a critical point or of a
first order phase transition. The first phase of this program has been
completed at RHIC~\cite{escan} with many results already available.

\begin{figure}[htb]
 \centering
 \resizebox{0.7\textwidth}{!}{\includegraphics{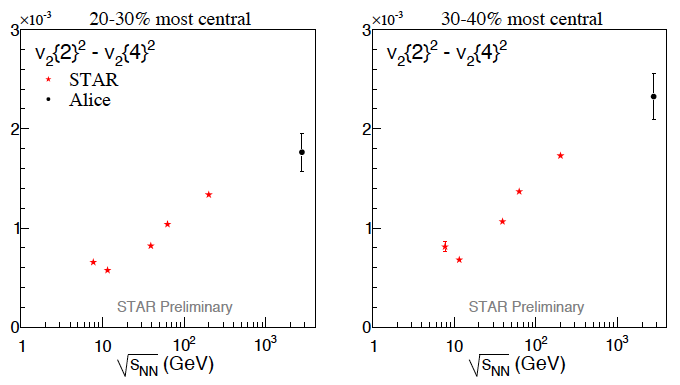}}
 \caption[]{ The width of the distribution of $v_2$ measured using the
   difference between the square of the second and fourth cumulants vs
   center of mass energy. Two centrality intervals are shown. }
 \label{fvfluc} \end{figure}
 
Besides looking for evidence of a critical point, which may or may not
lie in a region of the phase diagram accessible to heavy-ion
collisions, one can also look for indications of a softening of the
equation of state as the initial energy density decreases at lower
beam energies. Since the conversion of the initial state lumpiness
into the final state ridge-correlation and equivalently $v_n$
fluctuations is likely to be sensitive to the early pressure
gradients, we might expect these observables to be more sensitive to
changes in what part of the QCD equation of state the collisions are
probing. As the beam energy increases, the initial energy density
increases. If the pressure is increasing with energy density, then
collisions that start with higher energy density will have larger
pressure gradients. Those larger pressure gradients will be more
effective at converting the lumpiness of the initial energy density
into a lumpiness in momentum space. If collisions create initial
energy densities corresponding to the minimum in the pressure vs
energy density, then that lumpiness may tend to get washed out. This
is especially true given that the fluctuations are driven by hot-spots
in the initial overlap geometry that are approximately 1 fm in size.
Even a fairly brief period of free-streaming early in the evolution
can smear out hot-spots of that size.
If the lumpiness is washed out,
measurements of $v_3$ or $v_2$ fluctuations (accessed in data through
$v_2\{2\}^2-v_2\{4\}^2$) will show a minimum for collisions with
initial energy densities near where the pressure has a minimum.

Figure~\ref{fvfluc} shows $v_2\{2\}^2-v_2\{4\}^2$ as a function of the
colliding center of mass energy~\cite{qmvnrefs,mcm}. The preliminary
results do seem to indicate an increasing efficiency in the ability of
early pressure gradients to convert density fluctuations into $v_2$
fluctuations. This is exactly what we expect based on the Lattice QCD
calculations of the QCD equation of state (pressure vs. energy
density). In short, the picture that seems to emerge is that the $v_n$
fluctuations are a snapshot of the initial density distributions. The
clarity of that snapshot or the amplitude of the correlations, in
turn, provides a measurement of the pressure at the very earliest time
of the collision evolution. If the pressure is too low, the lumps
expand and smear out before they can be efficiently converted into
momentum space. Mapping out $v_n$ fluctuations vs beam energy is
therefore, a very exciting topic in heavy-ion collisions. Even if a
critical point is never found within the region of the phase diagram
that can be accessed by heavy-ion collisions (as perhaps appears
likely~\cite{fss}), experimentally observing the QCD equation of state
so directly, will be one of the most exciting results to emerge from
heavy ion physics.

\subsection{The Temperature of the Plasma}

In order to map out the regions of the phase diagram probed by
heavy-ion collisions and to study the thermodynamics of the matter
created in the collisions, it's important to estimate the temperature
of the QGP when it is first formed. There are several ways to do this
including by attempting to measure the color (or wavelength) of the
photons emitted from the source. This method requires disentangling
various sources of photons and then inferring the temperature from the
slope of the $p_T$ spectra of the thermal photons. Recent studies
indicate an inverse slope consistent with T=220 MeV~\cite{photons},
which is just above the expected transition temperature of 165 MeV.
The photon data is also consistent with hydrodynamic models which
start with temperatures in the range from 300 MeV to 600 MeV. This
suggests again, that indeed we are probing above the phase transition.

\begin{figure}[htb]
 \centering
 \resizebox{0.7\textwidth}{!}{\includegraphics{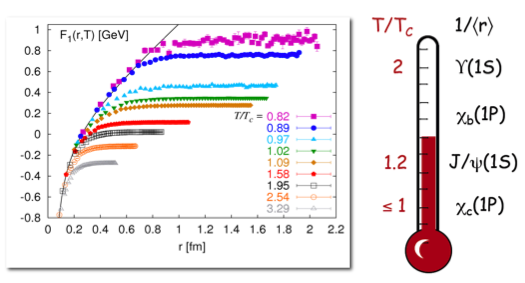}}
 \caption[]{ Quarkonium as a tool to measure the temperature. The
   left panel shows screening of the free energy for heavy quark and
   anti-quark pairs at finite temperature. The sequential melting of
   states can provide a thermometer for the
   QGP~\cite{Mocsy:2005qw,Mocsy:2007jz}.}
\label{f14}
\end{figure}
 
Another valuable tool for determining the temperature of the produced
matter is the Debye screening of heavy quark pairs (Quarkonium). The
idea is that in a hot and dense deconfined QGP, the abundant light
quarks can screen the force acting to bind the heavy quark pairs in a
quarkonium state~\cite{Matsui:1986dk}. The expectation of screening
has been confirmed in lattice QCD
calculations~\cite{Kaczmarek:2004gv,Petrov:2006pf}. Figure~\ref{f14}
shows the free-energy between two heavy quarks placed inside a bath of
light quarks. The zero temperature free energy is shown as a black
curve. As the temperature is increased the free energy drops further
and further from the zero temperature case. At first the deviation is
only at large distance, but as the temperature is increased, the
screening sets in at smaller and smaller distances. This means that
the large, loosely bound, quarkonium states will melt at low
temperatures while the more tightly bound states will survive until
hotter temperatures are reached. This sequential screening provides
and effective thermometer for determining the temperature of the QGP
by observing which Quarkonium states survive and which were unable to
form in the QGP~\cite{Karsch:1987pv,Digal:2001ue}. Measurements show
that the $J/\psi$ is suppressed at RHIC~\cite{Adare:2006ns}.
Measurements of higher excited states and the Upsilon still await more
precise measurements.

\subsection{The Opacity of the Plasma}

\begin{figure}[htb]
  \centering
  \resizebox{0.54\textwidth}{!}{\includegraphics{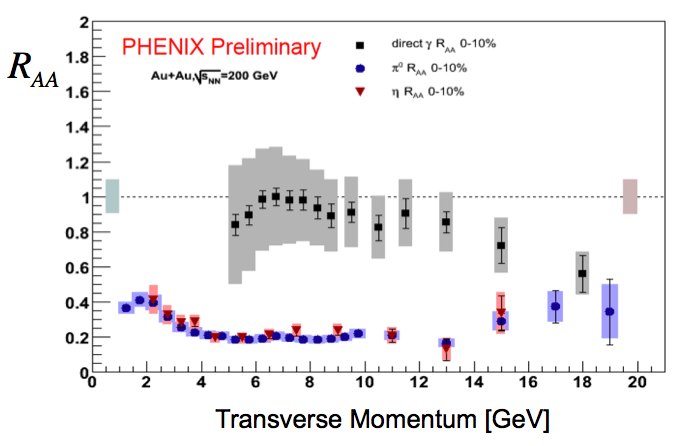}}
  \resizebox{0.44\textwidth}{!}{\includegraphics{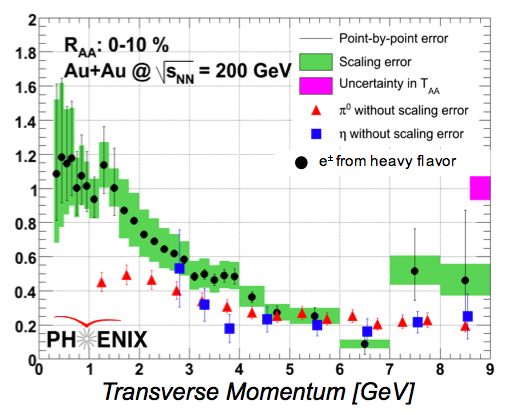}}
  \caption[]{ Suppression of high momentum particles do to the opacity
    of the QGP. The left panel shows that while high momentum pions
    and $\eta$-mesons are suppressed by a factor of 5, the color
    neutral photons are not.}
\label{f15}
\end{figure}

Another consequence of the formation of a hot and dense medium is jet
quenching~\cite{Wang:2000fq,Gyulassy:2001kr,Gyulassy:2000gk}. In a
heavy ion collision, the hard scattered partons that form jets must
traverse the hot and dense matter before they escape and hadronize in
the vacuum. The time-scales require that hard processes occur early in
the collision and the soft fragmentation process occurs late. As the
fast moving parton traverses the dense medium, it can lose energy to
the medium and be quenched. Since the number of high momentum
particles falls quickly, even a modest loss of energy can result in a
large suppression of high momentum hadrons observed in a particular
bin. Figure~\ref{f15} shows the ratio of particles observed in
heavy-ion collisions compared to what is expected based on scaling of
p+p collisions
($R_{AA}$)~\cite{Adler:2003qi,Adams:2003kv,Adler:2005ig,Adler:2006bv}.
The left panel shows that the number of pions and $\eta$-mesons with
momentum above 5 GeV is suppressed by nearly a factor of 5. The figure
also shows that photons in the same region are not suppressed, as is
expected since they are color neutral and do not lose energy in the
QGP.

It is also of quite some interests to note that, as seen in the right
panel of figure~\ref{f15}, even electrons coming from the decay of
heavy flavor hadrons are suppressed~\cite{Adare:2006nq}. This
indicates that the QGP is able to quench even high momentum
heavy-quark jets. Heavy quarks (\textit{e.g.} charm and bottom quarks)
where expected to lose less energy to the QGP
matter\cite{Dokshitzer:2001zm,Djordjevic:2003zk}. It's not a priori
obvious that heavy quarks will couple significantly to the medium and
be influenced by its apparent expansion. The extent to which they do
couple to the medium should be reflected in how large $v_2$ for heavy
flavor hadrons becomes and how much the nuclear modification
($R_{AA}$) deviates from unity. Precision measurements of Heavy Flavor
mesons or baryons are not yet available from the RHIC experiments. As
a proxy for identifying D-mesons, the STAR and PHENIX experiments have
measured non-photonic electrons\cite{Adler:2005xv,Abelev:2006db}.
Non-photonic electrons are generated from the weak-decays of heavy
flavor hadrons and after various backgrounds have been accounted for
can, with some caveats\cite{Sorensen:2005sm}, be used to infer the
$R_{AA}$ and $v_2$ of D-mesons.

The right panel of figure~\ref{f15} shows $R_{AA}$ for non-photonic
electrons\cite{Adler:2005ab,Adare:2006nq}. Prior to the measurement of
non-photonic electron $R_{AA}$, it was expected that heavy-flavor
hadrons would be significantly less suppressed than light flavor
hadrons. These expectations based on a decrease in the coupling of
charm quarks to the medium because of the dead-cone
effect\cite{Dokshitzer:2001zm}, are contradicted by the data; At $p_T
\approx 5$~GeV/$c$, non-photonic electrons are as suppressed as pions.
This suppression suggests a stronger than expected coupling of charm
quarks to the medium. This coupling apparently also leads to
significant $v_2$ for non-photonic electrons. The QGP created at RHIC
turns out to be exceptionally opaque, even to heavy-quark jets.

\begin{figure}[htb]
  \centering
  \resizebox{0.8\textwidth}{!}{\includegraphics{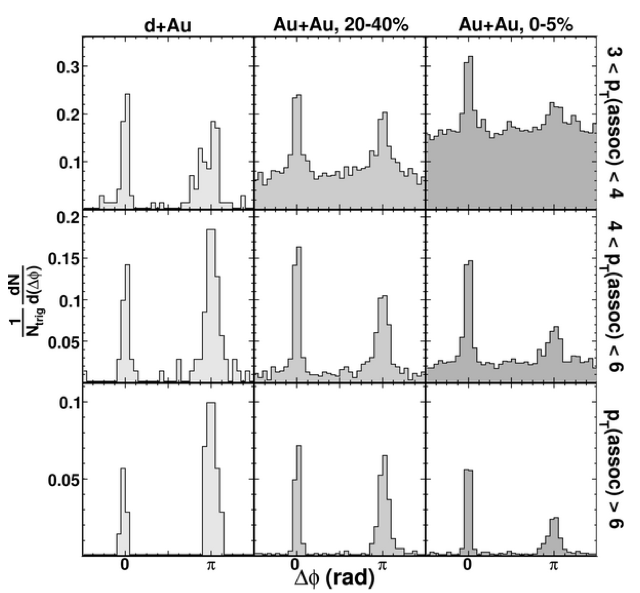}}
  \caption[]{ Jet-like correlations between the remaining high momentum hadrons. }
\label{f16}
\end{figure}

Evidence of the opacity of the QGP is also evident in the correlations
between high momentum particles~\cite{Adler:2002tq,Adams:2006yt}.
These measurements serve as a stand-in for jet studies which are very
difficult to carry out in heavy-ion collisions and are just now coming
to fruition. The two particle correlations however, already
demonstrate the opacity of the QGP to jets. The figure shows the
angular difference between an 8 GeV hadron and a lower momentum
associated hadron which satisfies the $p_T$ cut specified in the
figure. These kinematic selections seem to be sufficient hard to
resolve the correlations due to jet fragmentation as opposed to the
correlations at lower momentum dominated by flow and initial state
density fluctuations. The left panels of figure~\ref{f16} show the
results for d+Au collisions while the center and right panels show the
results for mid-central and central Au+Au respectively. The peak near
zero, is caused by the correlation between two particles coming from
the same parton. The peak at $\pi$ is caused by the correlation
between a hadron from a jet and a hadron from the away-side jet
created in the original scattering process. For d+Au collisions, the
away-side peak contains more pairs than the near-side peak. For
central Au+Au collisions the away-side peak is suppressed. This
suppression is consistent with an opaque medium. When looking for high
momentum particles, we have a preference for finding particles
originating from jets near the surface of the medium. When we look at
the away-side, we see the correlations from the jet that needed to
traverse the medium before escaping. The away-side jet is shown to be
quenched in the opaque medium.

\section{Conclusion}
Studies at RHIC have revealed that the collision of Au nuclei at
ultra-relativistic energies creates a hot and dense QGP that is opaque
to fast moving color charges, acts like a liquid, transforms into
hadronic matter through a smooth cross-over, and converts spatial
structure in the initial density distributions into correlations
between final particles in momentum space. Current studies are
searching for a first order phase transition at higher baryon chemical
potentials, a possible critical point, and/or indications relatively
rapid increases in the degrees-of-freedom indicating the onset of QGP.
Recent advances in understanding heavy-ion phenomenology and detector
upgrades will allow RHIC to remain at the forefront of nuclear physics
for another decade. The RHIC facility is ideally suited for these
studies and continues to provide fascinating insights on the nature of
QCD matter.

\end{document}